\documentclass[
 reprint,
superscriptaddress,
 amsmath,amssymb,
 aps,
]{revtex4-2}
\usepackage{float}
\usepackage{afterpage}
\usepackage{graphicx}
\usepackage{dcolumn}
\usepackage{xcolor}
\usepackage{bm}
\usepackage{graphicx} 
\usepackage{amsmath}
\usepackage{cancel}

\begin{document}

\preprint{APS/123-QED}

\title{Simple biological controllers drive the evolution of soft modes}

\author{Christopher Joel Russo}
\affiliation{Program in Biophysical Sciences, University of Chicago, Chicago, IL, USA}
\affiliation{James Franck Institute, University of Chicago, Chicago, IL, USA}
\author{Kabir Husain}
\affiliation{Department of Physics, University College London, London, United Kingdom}
\author{Rama Ranganathan}
\affiliation{The Pritzker School for Molecular Engineering, University of Chicago, Chicago, IL, USA}
\affiliation{Center for Physics of Evolving Systems, University of Chicago, Chicago, IL, USA}
\author{David Pincus}
\affiliation{Department of Molecular Genetics and Cell Biology, University of Chicago, Chicago, IL, USA}
\affiliation{Center for Physics of Evolving Systems, University of Chicago, Chicago, IL, USA}
\author{Arvind Murugan}
\affiliation{James Franck Institute, University of Chicago, Chicago, IL, USA}
\affiliation{Center for Physics of Evolving Systems, University of Chicago, Chicago, IL, USA}
\affiliation{Department of Physics, University of Chicago, Chicago, IL, USA}
\date{\today}


\begin{abstract}
Biological systems, with many interacting components, face high-dimensional environmental fluctuations, ranging from diverse nutrient deprivations to toxins, drugs, and physical stresses. Yet, many biological control mechanisms are `simple' -- they restore homeostasis through low-dimensional representations of the system's high-dimensional state. 
How do low-dimensional controllers maintain homeostasis in high-dimensional systems? We develop an analytically tractable model of integral feedback for complex systems in fluctuating environments. We find that selection for homeostasis leads to the emergence of a soft mode that provides the dimensionality reduction required for the functioning of simple controllers. 
Our theory predicts that simple controllers that buffer environmental perturbations (e.g., stress response pathways) will also buffer mutational perturbation, an equivalence we test using experimental data across ~5000 strains in the yeast knockout collection. We also predict, counterintuitively, that knocking out a simple controller will \emph{decrease} the dimensionality of the response to environmental change; we outline transcriptomics tests to validate this. Our work suggests an evolutionary origin of soft modes whose function is for dimensionality reduction in and of itself rather than direct function like allostery, with implications ranging from cryptic genetic variation to global epistasis.
\end{abstract}

\maketitle

How can a complex system be maintained at homeostasis without a controller that is as complex as the system itself?  Biological organisms have many possible internal degrees of freedom – even single celled organisms have thousands of genes that interact via complex regulatory networks \cite{milo2015cell}. These organisms are able to survive in face of diverse environmental fluctuations that perturb them in a high dimensional way; for example, environmental perturbations can involve different toxins, drugs and physical stresses like temperature, pH or osmolarity or the deprivation of different combinations of nutrients. 

Naively, homeostasis of such complex systems subject to high-dimensional fluctuations would require a complex controller. For example, a controller might need to monitor many specific aspects of the system state. This could include separately sensing the impacts of amino acid deprivation, carbon limitation, nitrogen limitation, and ribosome-targeting antibiotics, while taking distinct restorative actions for each. However, such a complex controller might be prohibitively costly for an organism to maintain. In contrast, many low-dimensional controllers have been documented across biology – these controllers project the high-dimensional state of the organism to much lower dimensional space and take restorative actions based on this projection (Fig. 1a). For example, in bacteria, while uncharged tRNAs for the distinct amino acids, carbon and phosphate deficiency, and temperature stress represent distinct types of environmental stress, they all converge to modulate the concentration of a single molecule, (p)ppGpp; this low-dimensional representation of the high-dimensional stresses then determines the `stringent response', regulating transcription, translation, replication, and other aspects of bacterial physiology \cite{wu2022cellular,traxler2008global,magnusson2005ppgpp} (Fig. 1c).  
\par 
Similarly, diverse aspects of the environment experienced by yeast -- glucose levels, amino acid abundance, nitrogen state, and osmotic stress -- are integrated into the level of cAMP \cite{conrad2014nutrient,thevelein1999novel} (Fig.1b). While distinct receptors are dedicated to sensing these different environmental perturbations, their downstream signals are projected down onto a low dimensional variable, the level of cAMP. The integrated cAMP variable then influences diverse downstream processes including transcriptional regulation, metabolic adjustments, and cell cycle control. In the innate immune system, the cAMP pathway homologously
senses diverse immune signaling molecules and controls complex immune
behavior, including phagocytosis and antimicrobial activity \cite{gottesman2019trouble,serezani2008cyclic}.

\begin{figure*}[!htb]
    \centering
    \includegraphics[width=\linewidth]{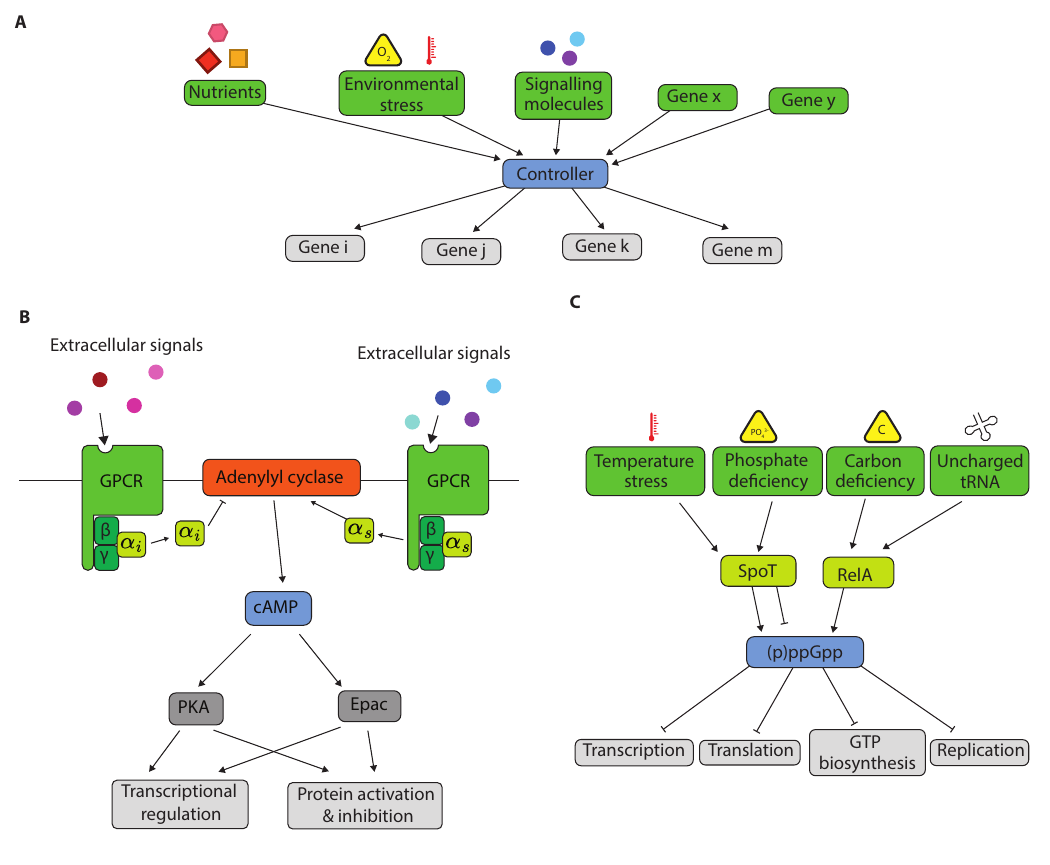}
    \caption{  (a) Low-dimensional controllers -- controllers that project the high-dimensional state of the organism and the environment to much lower dimensional space and take restorative actions based on this projection -- are found throughout biological systems.  (b) The cAMP pathway is a low-dimensional controller across eukaryotes: in yeast, diverse environmental inputs (glucose levels, amino acid abundance, nitrogen state, osmotic stress) are integrated into a single cAMP concentration, which then coordinates transcriptional regulation, metabolic adjustments, and cell cycle control (c) The stringent response is a prokaryotic low-dimensional controller: multiple stress signals (including carbon deficiency, phosphate deficiency, and temperature stress) are integrated into the concentration of (p)ppGpp, which accordingly reprograms the entire cellular physiology.}
    \label{fig1}
\end{figure*}
\par

Here, we build a theoretical framework to understand conditions where simple controllers control complex high-dimensional systems. We define 'simple' controllers as integral feedback mechanisms that regulate a high-dimensional system by first projecting its state onto a low-dimensional representation before sensing and acting. In contrast, 'complex' controllers directly monitor and respond to multiple independent aspects of the system's state without such dimensionality reduction.  Using simulations of gene regulatory networks and an analytically tractable model of integral feedback control, we show that selection for homeostasis leads to the evolution of a soft mode (alternatively called a dynamical slow mode) in the organism's dynamics if the controller is constrained to be `simple'. A soft or slow mode is a mode that relaxes slowly; when a system with a soft mode or modes is perturbed in the absence of a control mechanism, some components of the perturbation will decay quickly (stiff directions) while others decay much more slowly (soft directions).  That the mode is \textit{soft} means that arbitrary perturbations will tend to push the system along that mode \cite{bahar2010functional,leo2005analysis,russo2024soft,husain2020physical}. 
\par Our framework makes surprising predictions that we test with empirical data.  Our model predicts that proteins mediating environmental robustness, such as stress response factors, also facilitate mutational robustness, a prediction we test using data from ~5000 strains in a yeast knockout collection \cite{costanzo2016global,costanzo2021environmental}. Our framework predicts that knocking out a controller will \emph{reduce} the dimensionality of environmental perturbation effects, a prediction supported by yeast transcriptomic data, with further experiments suggested.
\par
Prior works have presented explanations that invoke soft modes for the widely observed low dimensional nature of biological systems. These works predict soft modes with a specific structure -- the soft mode is assumed to be directly functional, including for communication through allostery in proteins  \cite{mitchell2016strain,rocks2017designing,campitelli2020role,ravasio2019mechanics,raman2016origins} or as internal models of low-dimensional structure in the external world \cite{friedlander2015evolution,furusawa2018formation}.  In contrast, in our framework, the soft mode can point in any direction in state space, as we do not assign any specific function to the soft mode itself. Instead, the presence of any soft mode leads to dimensionality reduction that can be exploited by simple feedback controllers. Large-scale datasets increasingly reveal low-dimensional structure in biological systems across different scales. Our work provides a scale-independent mechanistic explanation for this phenomenon while offering testable predictions for future studies examining both environmental and genetic variation.

\section*{Results} 

\begin{figure*}[!htb]
    \centering
    \includegraphics[width=\linewidth]{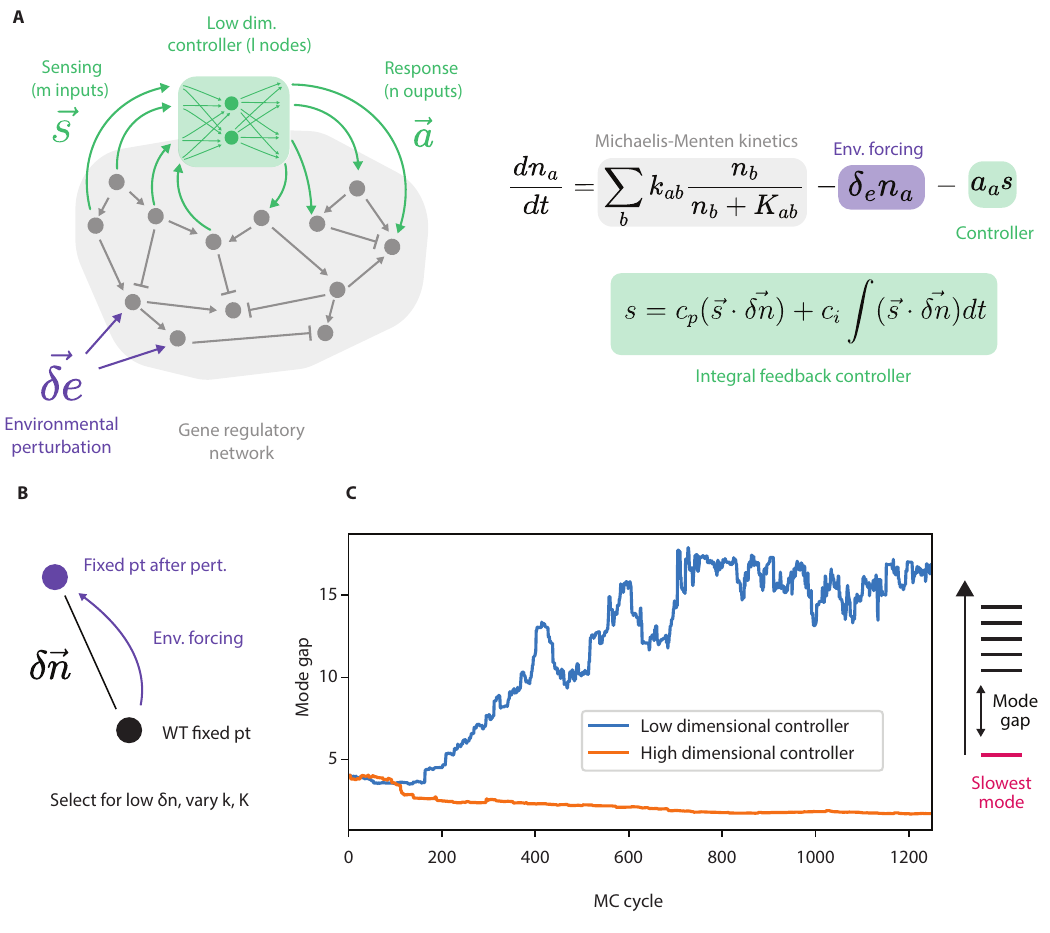}
    \caption{Selection for environmental robustness in a system with a low complexity controller leads to the evolution of a dynamical slow mode. (a) We simulate a simple model of a gene regulatory network. In gray, Michaelis-Menten kinetics govern interactions between genes, while the controller mechanism and environmental forcing terms are highlighted in green and purple, respectively. The controller measures $m$ nodes, projects them down to $k$ random and independent combinations, and then acts on m nodes based on that project. $k = 1$ resembles the simple controllers found in biology while large $k$ correspond to more complex controllers.  (b) Selection for robust networks, quantified by the normalized average deviation from the WT fixed point in the presence of a low complexity controller (c) leads to an increase in the mode gap of the system (blue). Selection for robustness in the presence of a high complexity controller does not lead to an increase in mode gap (orange). }
    \label{fig2}
\end{figure*}
\subsection{In-Silico Evolution Simulations}
How are simple controllers able to regulate high dimensional biological systems in the face of diverse challenges? To address this question, we consider a model of a complex gene regulatory network that has a stable point and can be disturbed by high-dimensional environmental perturbations acting on different network nodes (Fig. 2a):
\begin{equation}
    \frac{dn_a}{dt} = \sum_b k_{ab}\frac{n_b}{n_b + K_{ab}} + \delta e_a(t) n_a
\end{equation}
where $n_a$ is the expression level of gene $a$ and $k_{ab}$ and $K_{ab}$ define the interactions between genes $a$ and $b$, $e_a$ represents environmental fluctuations that act on different genes $a$ of the network. 

We now add an integral feedback controller that tries to restore homeostasis -- to return the system to its fixed point $x_0$, despite environmental fluctuations $\vec{e}(t)$. Integral feedback control mechanisms are widely implemented at the cellular level in organisms for maintaining homeostasis and responding robustly to environmental conditions \cite{yi2000robust,somvanshi2015implementation}. 

Integral feedback is often formulated for a single variable $n(t)$ and is based on computing an integrated error signal,
\begin{equation}
    I(t)  = c_i \int^{t} \delta n(t') dt' + c_p \delta n(t)  
\end{equation}
where $\delta n(t) = n(t) - n_0$ is the error signal, i.e., deviation from the fixed point $n_0$ (Fig.2b); $c_i, c_p$ are the gain parameters for the integral and proportional parts of the feedback control. This integrated error signal is then used to take restorative action for $n(t)$, e.g., $dn/dt = f(n) - I(t)$.

For our high-dimensional system, we consider an integral feedback controller that measures $k$ independent integrated error signals $I_i, i = 1, \ldots k$, each of which measures the deviation of a different linear combination $\vec{s}_i$ $\vec{s}_i \cdot \vec{n} = \sum_a s_i^{a} n_a$ of the nodes $n_a$ from their set points,
\begin{equation}
    I_i(t) = c_p(\vec{s}_i \cdot \vec{\delta n}) + c_i \int (\vec{s}_i \cdot \vec{\delta n}) dt.
\end{equation}
When the number of measurements ($k$) equals the total number of nodes ($N$), the controller can monitor the complete state of the system, sensing $N$ combinations $\vec{s}_i \cdot \vec{\delta n}$ of deviations. However, if $k<N$, the controller must work with incomplete information, monitoring only a simplified version of the system's full state, a $k$-dimensional projection of the $N$ dimensional state of the system. Intuitively, $k$ can be understood as the internal complexity of the controller as shown in Fig. 2a. The simplest case, $k = 1$, represents a controller like those often seen in biological systems - where diverse signals are compressed into a single measure like cAMP or ppGpp levels. 

\par To understand how controllers with low $k$ values could ever be effective, we turn to in silico evolution of genetic networks under fluctuating environmental conditions (Fig.2a). We constrain the controller to either low ($k=1$) or high ($k= 30$) complexity and evolved parameters of the underlying dynamical system (here, $k_{ab}$ and $K_{ab}$)

\begin{equation}
    \frac{dn_a}{dt} = \sum_b k_{ab}\frac{n_b}{n_b + K_{ab}} + \delta e_a(t) n_a - \sum_{i=1}^k g_a I_i 
\end{equation}

We add a controller that tries to restore the system to the wild type fixed point in the face of environmental challenges:
\begin{equation}
    \frac{dn_a}{dt} = \sum_b k_{ab}\frac{n_b}{n_b + K_{ab}} -g_a s + \delta f_a n_a
\end{equation}

\begin{equation}
    s = c_p(\vec{s}\cdot \vec{\delta n}) + c_i \int (\vec{s}\cdot \vec{\delta n}) dt
\end{equation}
where $\vec{s},\vec{g}$ are parameters that determine the behavior of the controller and $\vec{\delta n}$ is $\vec{n}_{WT}-\vec{n}$, the deviation from the WT fixed point (Fig.  2b). We define the \textit{fitness cost} of the system as the failure to maintain homeostasis, quantified by residual deviation from the fixed point a low absolute value of $\vec{\delta n}$, normalized by the overall stiffness of the system, averaged over many different values of environmental fluctuations $\vec{\delta f}$. We simulate the evolution of the network, selection for higher fitness using a simulated annealing Markov Chain Monte Carlo protocol (see supplement).
\par 
We find strikingly different results for low and high $k$: constraining controller complexity (low $k$) leads to dimensionality reduction in the controlled system itself while high $k$ controllers allow the system to retain a high dimensional response to perturbations (Fig. 2c). Selection for robustness with a $k=1$ controller leads to a reduction in the system's \textit{mode gap} - the ratio of the second and first eigenmodes of the system around its fixed point, $\frac{\lambda_1}{\lambda_0}$, which corresponds to the emergence of a soft mode. We show that selection for robustness given a simple controller leads to the development of a slow mode that channels perturbations. A slow mode \cite{russo2024soft} is also a soft mode -- when arbitrarily perturbed, the system tends to move along this mode, making it easier for a simple controller to detect and correct deviations. 

\subsection{Analytic Theory for a Low Complexity Controller}

Here we use a simple analytically tractable model of integral feedback control to build intuition about the fitness value of a mode gap and the resulting soft mode. Consider a system with dynamics $f$ in the absence of any control but is subject to environmental perturbations

\begin{equation}
    \frac{d \vec{x}}{dt} = \vec{f}(\vec{x}) + \vec{e}_{env}
\end{equation}
where $\vec{e}_{env}$ represents high-dimensional environmental perturbations to the state of the system $\vec{x}$. 

In such a system, the impact $\delta \vec{x}$ of environment perturbation on the state of the system will be of the form,
\begin{equation}
    \delta \vec{x} = J_x^{-1} \delta \vec{e}_{env}
\end{equation}
where $J_x = \{ \frac{\partial f_i}{\partial x_j } \}$ 
is the Jacobian with respect to the state $x$ of the system. 

The distribution of eigenvalues $\{\lambda_0,\lambda_1...\lambda_n\}$ of $J_x$ affects the behavior of the system in response to an environmental perturbations. 

We find that the impact of environmental perturbations $\delta \vec{x}$ will be as high dimensional as the environmental perturbations $\delta \vec{e}_{env}$ in the absence of a mode gap but can be much lower dimensional with a mode gap. 

Concretely, if we consider an ensemble of environmental perturbations $\{\vec{\delta e}_{env}\}$, we can quantify its effective dimensionality using effective rank\cite{roy2007effective}:
\begin{equation}
    \text{rank}_{\text{eff}}(\{e_{env}\}) = \exp \Big(-\sum p_i \log(p_i)\Big)
\end{equation}
where $p_i$ is the fraction of variance explained by the $i$th principal component. 

We can in the same way compute the effective rank of the ensemble of the \textbf{impacts} of these environmental perturbations $\text{rank}_{\text{eff}}(\{\vec{\delta x}\})$
We can show that it has the same effective rank as the ensemble of the environmental perturbations that caused them:
$$ \mbox{rank}_{\text{eff}}(\{\vec{\delta x}\})  \approx \mbox{rank}_{\text{eff}}(\{\vec{\delta e}_{env}\}) $$
with the assumption that all eigenvalues of $J_x$ are the same value, $\lambda$ (Fig. 3a).

On the other hand, if the eigenvalues show a mode gap, (i.e. $\frac{\lambda_1}{\lambda_0} >>1$), the effective rank $ \mbox{rank}_{\text{eff}}(\{\vec{\delta x}\})$ can be \textit{lower} than that of $\mbox{rank}_{\text{eff}}(\{\vec{\delta e}_{env}\})$.  If we consider a system with a single slow mode $\lambda_0$ and all other modes $\lambda$, with a mode gap $\frac{\lambda}{\lambda_0}$ and a ensemble of perturbations where each entry $\delta e_{env,i} \sim N(0,\sigma)$
\begin{equation}
    \mbox{rank}_{\text{eff}}(\{\vec{\delta e}_{env}\})  >> \mbox{rank}_{\text{eff}}(\{\vec{\delta x}\})  
\end{equation}
The effective rank $\mbox{rank}_{\text{eff}}(\{\vec{\delta x}\})$ is greatly reduced from $\mbox{rank}_{\text{eff}}(\{\vec{\delta e}_{env}\})$ ( see supplement for full derivation). 

Intuitively, the effective rank (or dimensionality) of $\{\vec{\delta x}\}$, the impact of environmental perturbations, can be much lower dimensional than the environmental perturbations $\{\vec{\delta e}_{env}\}$ if the system's Jacobian $J_x$ has a slow mode that dominates the effect of perturbations (Fig. 3b). 
\par

We now show that integral feedback can exploit this reduced dimensionality of impacts $\delta x$ to function effectively despite the high dimensional nature of environmental fluctuations $e_{env}$.
 We consider integral feedback of low complexity, i.e., a single sensing vector $\vec{s}$ compresses information about the state of the system into a one dimensional scalar (Fig. 3e). Integral feedback tries to restore the system based on this scalar $s$ by acting along an action vector $\vec{a}$, determining the direction the integral feedback restores along. $k_i$ and $k_p$ are the integral and proportion terms, respectively.
\begin{equation}
    \frac{d \vec{x}}{dt} = f(\vec{x}) +\vec{\delta e}_{env}- \Vec{a}\Big[ k_p (\Vec{s}\cdot \vec{\delta x}) + k_i \int (\Vec{s}\cdot \vec{\delta x}) dt \Big]
\end{equation}
We consider the linear regime about the fixed point. 
Solving for the impact vector $\vec{\delta x}$ in the presence of a controller, we find,
\begin{equation}
    \vec{\delta x} =  \tilde{\delta e} -\Bigg(\frac{ \tilde{\delta e}\cdot\vec{s} }{\tilde{ a} \cdot \vec{s}  } \Bigg)\tilde{a}
\end{equation}
with rescalings $\tilde{\delta e}_i = \delta e_{env,i}/\lambda_i$, and $\tilde{\delta a}_i = \delta a_i /\lambda_i $.

\par
As before, we consider a system with one slow mode $\lambda_0$ and other modes of faster and equal $\lambda$. We consider sensing $\vec{s}$ and action $\vec{a}$ unit vectors that sense and act on a direction that is a linear combination of the slow mode itself $\hat{v_0}$ and a unit vector in the direction of the mean perturbation $\hat{\mu}$: $\vec{a} = \beta \hat{v_0} + \sqrt{1-\beta^2}\hat{\mu}$ , $\vec{s} = \alpha \hat{v_0} + \sqrt{1-\alpha^2}\hat{\mu}$.
We find the expectation value $\langle||\vec{\delta x}||^2\rangle$ over a normally distributed $\vec{\delta e}$ as a function of  mode gap $ \lambda/\lambda_0$, the variance and mean of of the normal distribution $\vec{\sigma}_{env}$ and $\vec{\mu}_{env}$ and the parameters $\alpha, \beta$. We find that in the limit of a large mode gap $\lambda/\lambda >> 1$, the optimal values of  $\alpha, \beta$ are both 1. As the mode gap $\lambda/\lambda_0$ increases, the optimal $\vec{a},\vec{s}$ vectors (that is, those that minimize $\langle||\vec{\delta x}||\rangle$) align with the slow mode $\hat{v_0}$(see the supplement for a full derivation). With optimal controller parameters $\alpha,\beta$, in this limit the expectation value of the effect of a perturbation reduces to
\begin{equation}
    \langle||\vec{\delta x}||\rangle = \frac{1}{\lambda}\langle||\vec{\delta e}_{\perp\hat{v_0} }||\rangle = \frac{1}{\lambda}\sqrt{\sum_{i\neq 0} (\mu_i^2 + \sigma^2_i)}
\end{equation}
The soft mode is completely canceled, and all that is left is the effect on the uncontrolled fast modes. The expectation value $\langle||\vec{\delta x}||\rangle$ with optimal $\vec{a},\vec{s}$ normalized by the effect of an uncontrolled perturbation decreases as mode gap increases -- a higher mode gap leads to a more robust system, given a low dimensional controller (Fig. 3e).
\begin{equation}
    \langle||\vec{\delta x}||\rangle_{norm} = \frac{ \sqrt{\sum_{i\neq 0} (\mu_i^2 + \sigma^2_i)}}{\sqrt{\sum_{i\neq 0} (\mu_i^2 + \sigma^2_i) + (\lambda/\lambda_0)^2(\mu_0^2 + \sigma^2_0) }}
\end{equation}The fitness benefit of a high mode gap -- $\langle||\vec{\delta x}||\rangle_{\lambda/\lambda_0=1} - \langle||\vec{\delta x}||\rangle_{\lambda/\lambda_0=100}$ -- increases with the overall dimensionality of the system (Fig. 3g, see supplement for full derivation). Our simple analytic model above shows how slow modes can reduce the effective dimensionality of the effect of perturbations on a system, and in turn how simple 
\begin{figure*}[p!]
    \centering
    \includegraphics[width=\linewidth]{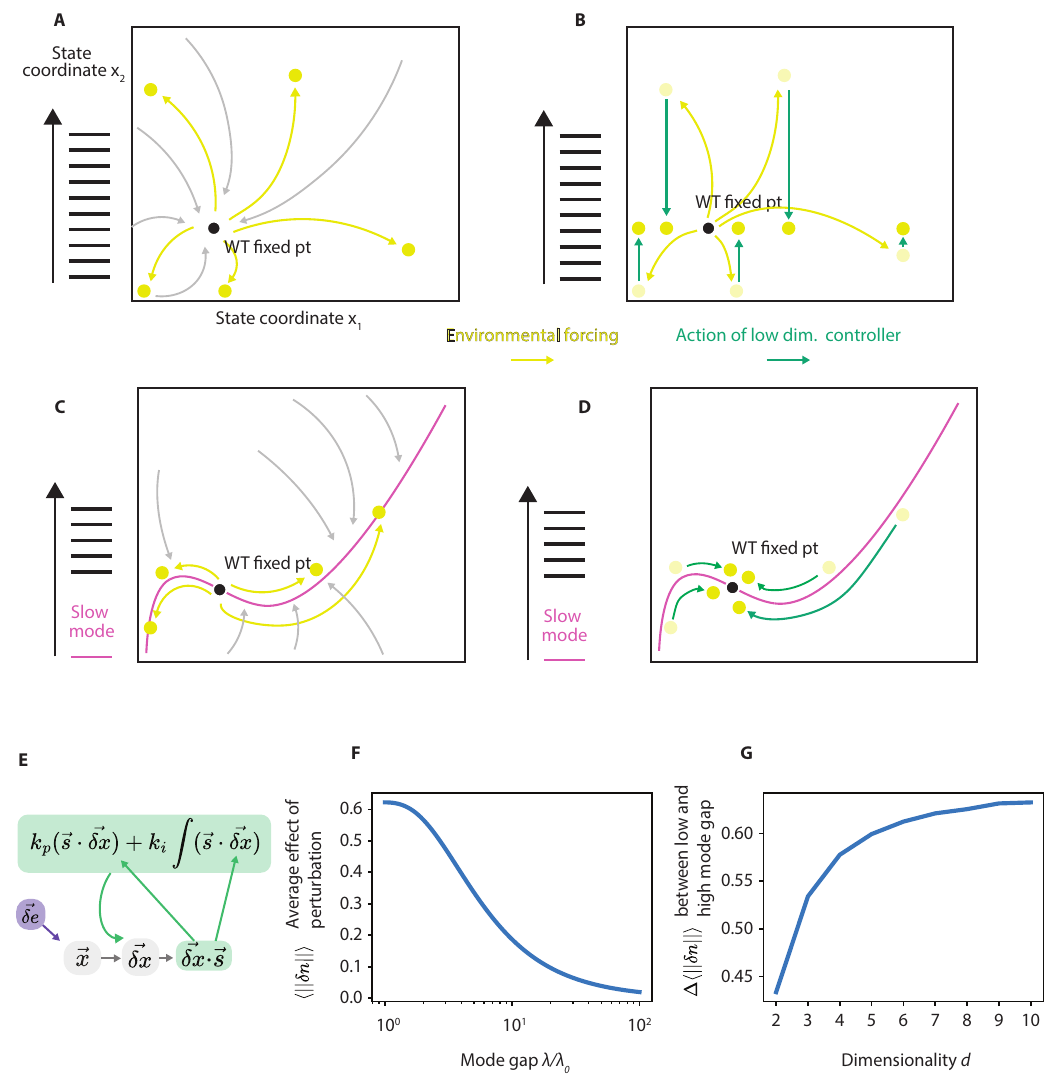}
    \caption{(a) In systems with no significant mode gap, relaxation trajectories are diverse, and perturbations push the system in arbitrary directions. (b) A low complexity controller is not effective in the face of diverse perturbations. (c) In systems with a significant mode gap, a quick relaxation to the slow mode is followed by slow relaxation along the slow mode. Environmental and mutational perturbations push the system largely along the slow mode. (d) A simple controller can nevertheless be effective because all it needs to do is sense and act on the slow mode. (e) In a linear model of a simple controller with integral feedback, as mode gap increases, (f) As mode gap increases, the normalized average effect of a perturbation $\langle ||\delta x||\rangle$ decreases. (g) As dimensionality of the system increases, the difference in normalized $\langle ||\delta x||\rangle$ between a system with a small mode gap and a large mode gap increases.}
    \label{fig3}
\end{figure*}
controllers can thus select for slow modes to facilitate robustness.

\subsection{Environmental buffers are mutational buffers}
\subsubsection{Theory}
Our model of slow mode mediated response to environmental challenges in living systems predicts dual buffering – mechanisms that buffer environmental perturbations should also buffer mutational perturbations. This dual ability to buffer environmental stressors as well as mutations has been noted since the 1990s in the case of the heat shock system in Drosophila \cite{rutherford1998hsp90,queitsch2002hsp90}. The slow mode is also a soft mode, and any perturbation tends to push the system along that mode \cite{queitsch2002hsp90,yeyati2008incapacitating}. In a simple linear model, the change in state due to environmental perturbation $\delta \vec{e}$ is $\delta_e \vec{x} \approx \sum_i \frac{\textbf{J}_e \delta \vec{ e} \cdot \vec{v_i}}{\lambda_i} \vec{v}_i$. The change to the cell state due to mutation $\delta \vec{g}$ is $\delta_g \vec{x} \approx \sum_i \frac{\textbf{J}_g \delta \vec{ g} \cdot \vec{v_i}}{\lambda_i} \vec{v}_i$.
If $\lambda_1 >> \lambda_{i>1}$ then 
\begin{equation}
    \delta_g \vec{x} \approx  \frac{\textbf{J}_g \delta \vec{ g} \cdot \vec{v_1}}{\lambda_1} \vec{v}_1
\end{equation}
and
\begin{equation}
    \delta_e \vec{x} \approx  \frac{\textbf{J}_e \delta \vec{ e} \cdot \vec{v_1}}{\lambda_1} \vec{v}_1
\end{equation}
Both kinds of perturbations will tend to be aligned. If a controller restores homeostasis by sensing the position of the system and restoring it along this mode, it will thus be able to buffer both kinds of challenges (Fig. 4a).
\par
	We use our simulations from above to explore dual buffering. We simulate the effect of mutational perturbations by introducing small changes to the interaction parameters of our Michaelis-Menten network, and test how low complexity controllers buffer the effects of mutations. We reuse controller parameters selected for environmental robustness, and we investigate genetic networks with and without slow modes. When there is a large mode gap, a controller that has evolved to buffer the effects of mutational forcing is able to buffer the effects of mutations as well (Fig. 4b). A low complexity controller that has been selected for environmental buffering in the absence of a mode gap is not able to buffer mutations. Our model of stress response and low dimensionality thus makes a prediction testable in perturbative experiments.

\subsubsection{Empirical Results}
Two experimental studies collected high throughput fitness data on yeast grown subject to different kinds of mutational and environmental perturbations, shedding light on dual buffering in living cells \cite{costanzo2016global,costanzo2021environmental}. A library of yeast strains with all non-essential gene knockouts had previously been assembled \cite{costanzo2010genetic}. In one experiment, this library was used to create a library of all possible pairwise double knockouts, which we term GxG (essential gene knockdowns were also assayed, as well as a subset of triple gene knockouts and knockdowns, but we do not analyze them in this work) \cite{costanzo2016global}. The fitness of the yeast strains in the GxG library was then assayed using an automated colony size protocol (Fig. 4c). In another experiment by the same group, each strain from the same single gene knockout library was subject to 14 diverse environmental perturbations (GxE). The fitness of each strain-environment combination was then assayed using the same colony size quantification protocol \cite{costanzo2021environmental}. In these experiments, the fitnesses of the single knockouts (G) and WT subject to environmental stresses (E) were also quantified using the same protocol. 
\par
To identify genes that buffer environmental or mutational perturbations, we first identify pairs of knockouts (in the GxG experiment) or knockout-environment pairs (in the GxE experiment) where there is a significant negative non-additive effect of the pair of perturbations (beyond the independent effects of the knockouts or the knockout and the environment, known from the G and E datasets). We use the same significance test as the authors of the original studies for identifying significant negative interactions (see supplement). In the GxG experiment, this is equivalent to identifying significant negative epistatic interactions. Genes which have many negative epistatic interactions with other genes are buffering many genes - the absence of the buffering gene increases the fitness cost of many other mutations. Similarly, genes with many significant negative interactions with environmental conditions buffer the effect of the environments – the absence of the gene increases the fitness cost of many environments.
\par
We find that genes that buffer many environmental conditions are more likely to buffer the effects of many mutational perturbations. We show this with a plot of histograms of genes by the number of knockouts the buffer, binned by the number of environmental conditions they buffer (Fig. 4d). Genes that buffer 6 or 7 environmental conditions are much more likely to buffer the effects of 300 or more genes than genes that buffer only 1 or 2 environmental conditions. One of the environmental challenges was subjecting the cells to Rapmycin, which is an mTOR inhibitor \cite{ballou2008rapamycin}. If we treat this as equivalent to an mTOR knockdown and apply the same analysis as above, we find that mTOR is buffering over the effect of over 400 knockouts. 

\par
These experiments show that genes that are able to buffer many environmental challenges are more likely to buffer many mutational challenges, in line with our theoretical predictions. 
\begin{figure*}[!htb]
    \centering
    
    \includegraphics[width=\linewidth]{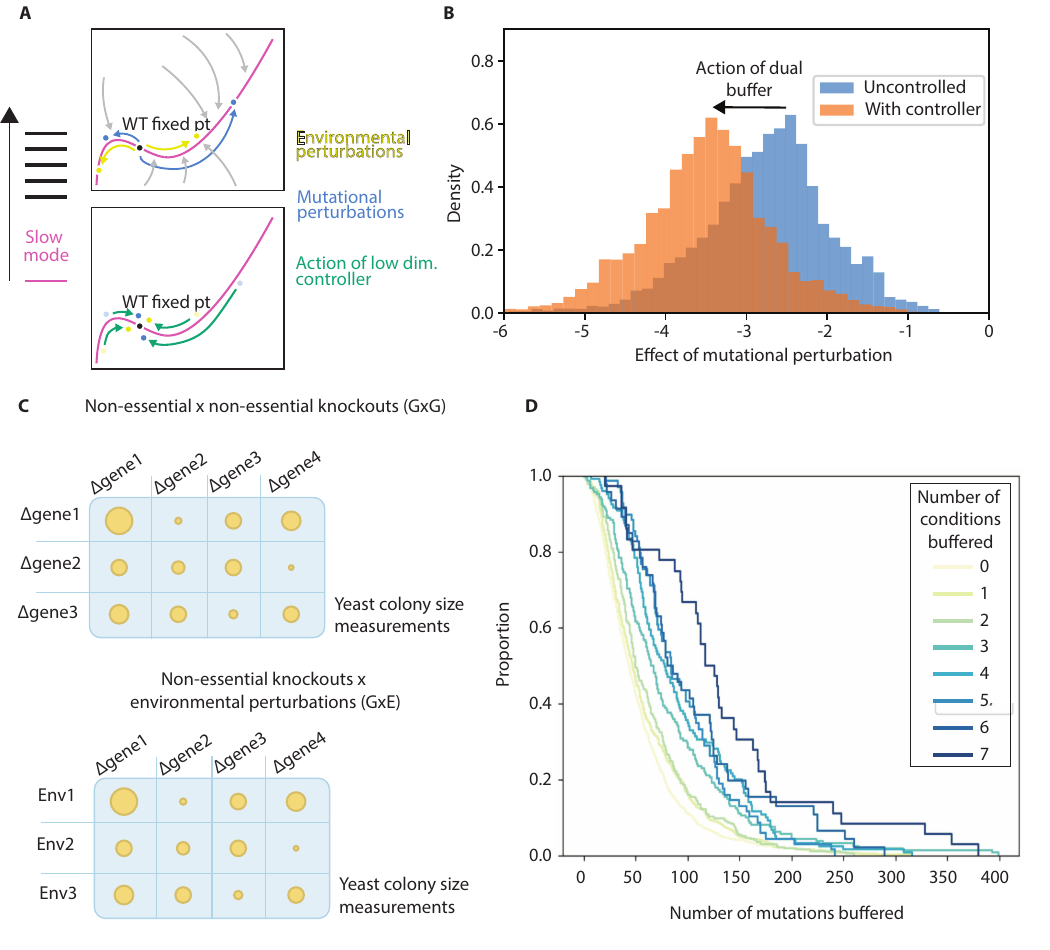}
     \caption{Slow modes allow controllers of environmental perturbations to also buffer mutational perturbations (dual buffering). (a) Slow modes channel all perturbations, regardless if they are environmental or mutational, along that mode; accordingly, a controller that senses and restores the state of the system along that mode is effective against both kinds of challenges  (b) Our simulations demonstrate that low complexity controllers,with controller parameters selected for environmental robustness, can buffer mutational perturbations, but only in networks with a large mode gap. (c) Two experimental studies measured yeast fitness: one using a library of all possible pairwise non-essential gene knockouts (GxG) \cite{costanzo2016global}, and another exposing single gene knockout strains to 14 diverse environmental perturbations (GxE) \cite{costanzo2021environmental}, both quantified via automated colony size measurements. (d) Significant GxE and GxG interactions were identified from the dataset. Genes are binned by the number of significant negative GxE interactions, and the cdf histograms of significant negative GxG interactions are plotted. Genes buffering multiple environmental conditions are significantly more likely to buffer numerous mutational perturbations, with genes buffering 6-7 environmental conditions much more likely to buffer 300+ gene knockouts than those buffering only 1-2 conditions, supporting our theoretical predictions of dual buffering.
 .}
    \label{fig5}
\end{figure*}

\subsection{Dimensionality increase by controllers}

 We simulate the effects of diverse environmental perturbations in our system with and without the controller knocked out, using our coevolved controller and slow-mode containing network. We apply many high dimensional environmental perturbations to the system, and then perform PCA on the resulting states (Fig. 5a). We show that the effective dimensionality is in fact lower once the controller is knocked out – the first principal component captures a much larger share of the variation caused by environmental challenges in the absence of the controller. In our model, the system evolves a slow mode, but because the mode is effectively buffered by a controller, the controlled system does not have an exposed soft mode. If the controller that listens to and restores homeostasis along the slow mode is knocked out, however, the knockout will expose the soft mode (Fig. 5b). This has the counterintuitive effect of reducing the dimensionality of the effects of environmental perturbations. Since the slow mode is soft, but uncontrolled with the controller knocked out, environmental stresses will push the system along that mode significnoantly in the absence of the controller.
 
\subsubsection*{Empirical results}
In another experimental study, the effects of kinase inhibition and environmental stresses on gene expression in yeast were investigated \cite{mace2020multi}. A panel of 32 mutant kinase strains, which allow for selective inhibition of the mutant kinase, along with wild-type replicates were grown to exponential phase in rich media (Fig. 5c). The 32 yeast strains were subjected to 10 different environmental conditions: rich media (YPD), synthetic media (SDC), heat shock (39°C), hyperosmotic shock (0.5 M NaCl), glucose depletion, endoplasmic reticulum stress (tunicamycin), oxidative stress (menadione), proteotoxic stress (AZC), TOR inhibition (rapamycin), and antifungal drug exposure (fluconazole). After 20 minutes of exposure to each condition, cells were harvested and bulk transcriptomic datasets were collected. In this work, we will focus on a comparison of Tpk123 inhibition with the WT controls. Tpk123 mediates the cyclic AMP system in yeast, a system which is known to have an hourglass topology as shown above, and is a possible low-dimensional controller.
\par
To investigate the effect of knocking out a low dimensional controller, we analyze the processed data from the original study – DESeq2 had been used to generate normalized expression values \cite{love2014differential}, and then the log2 fold change of each gene in each sample with respect to wild type cells in YPD was calculated. The original study computed t-distributed stochastic neighbor embedding (t-SNE) \cite{van2008visualizing} analysis (following PCA) of the data from the original study. 
\par
We find that diverse environmental challenges lead to a stereotyped transcriptional response in the context of Tpk123 inhibition. The t-SNE analysis shows that while the diverse environmental challenges tend to push the transcriptional state of the yeast cells in idiosyncratic ways in the absence of Tp123 inhibition (even when other kinases are inhibited), Tpk123 inhibited cells cluster together, across diverse perturbations (Fig. 5d,e,f). 
\par
If indeed a simple controller is buffering environmental challenges by listening to and restoring the cell state along a soft mode, inhibiting the controller will expose a soft mode. Diverse challenges in the absence of the controller will tend to push the system along the soft mode, and the observed dimensionality of the state of the cell will be lower across perturbations than in the presence of the controller. 
\section*{Discussion}

Low effective dimensionality in naively high dimensional biological systems has been observed in many different contexts across scales \cite{leo2005analysis,eckmann2021dimensional,Raimondi2010,sastry2019escherichia,sastry2019escherichia,schmidt2016quantitative,lukk2010global,kinsler2020fitness}.

It has been argued that in many contexts, from proteins to ecosystems, slow modes may mediate this low dimensionality \cite{leo2005analysis,husain2020physical,ravasio2019mechanics}. Many systems seem to have a slow mode to which dynamics quickly converge. But generic reasons for the emergence of such dimensionality reduction has been actively sought by the field \cite{russo2024soft, kashtan2005spontaneous,kaneko2024constructing}. 
\par
    
    We show that constraining controllers to be simple leads to the evolution of slow modes.
   
    In a linear model of proportional-integral feedback control, we show that a larger mode gaps facilitate the fitness of systems regulated by a low dimensional controller, with the controllers listening to and disproportionately acting on the slow mode.
    
    In a non-linear model of a high dimensional genetic network with Michaelis-Menten kinetics and a simple integral feedback controllers, we find that selection for robustness leads to the formation of a dominant slow mode.
    \par Others have proposed other direct function reasons for slow modes allostery. Selection for allosteric communication between distal parts of a network can lead to slow modes \cite{ravasio2019mechanics,rocks2017designing}. Systems which require a low dimensional internal model of the external world may do so via a slow mode within a system \cite{friedlander2015evolution,furusawa2018formation}. In prior work on  modularity, the dimensionality of the external pertubations was low dimensional, which ends up reflected in the internal dimensionality \cite{friedlander2015evolution, kashtan2005spontaneous}. In contrast, in our model the external world needs to be high dimensional, and the system itself evolves to make the internal effect low dimensional, so a simple controller can be effective.
Yet others have argued that evolvability, not direct function, has led to the evolution of slow modes \cite{raman2016origins, halabi2009protein,furusawa2018formation,masel2006cryptic}.

 \par In this work, we so far presented one direct function explanation for the evolutionary origin of such slow modes. However, the fact of the predicted duality means that in fact selection for mutational robustness (an evolvability benefit) could lead to slow modes that in turn facilitate environmental robustness. It may in fact not be possible to distinguish between these, since one leads to the others in the context of slow modes.
 \begin{figure*}[p!]
    \centering
    \includegraphics[width=\linewidth]{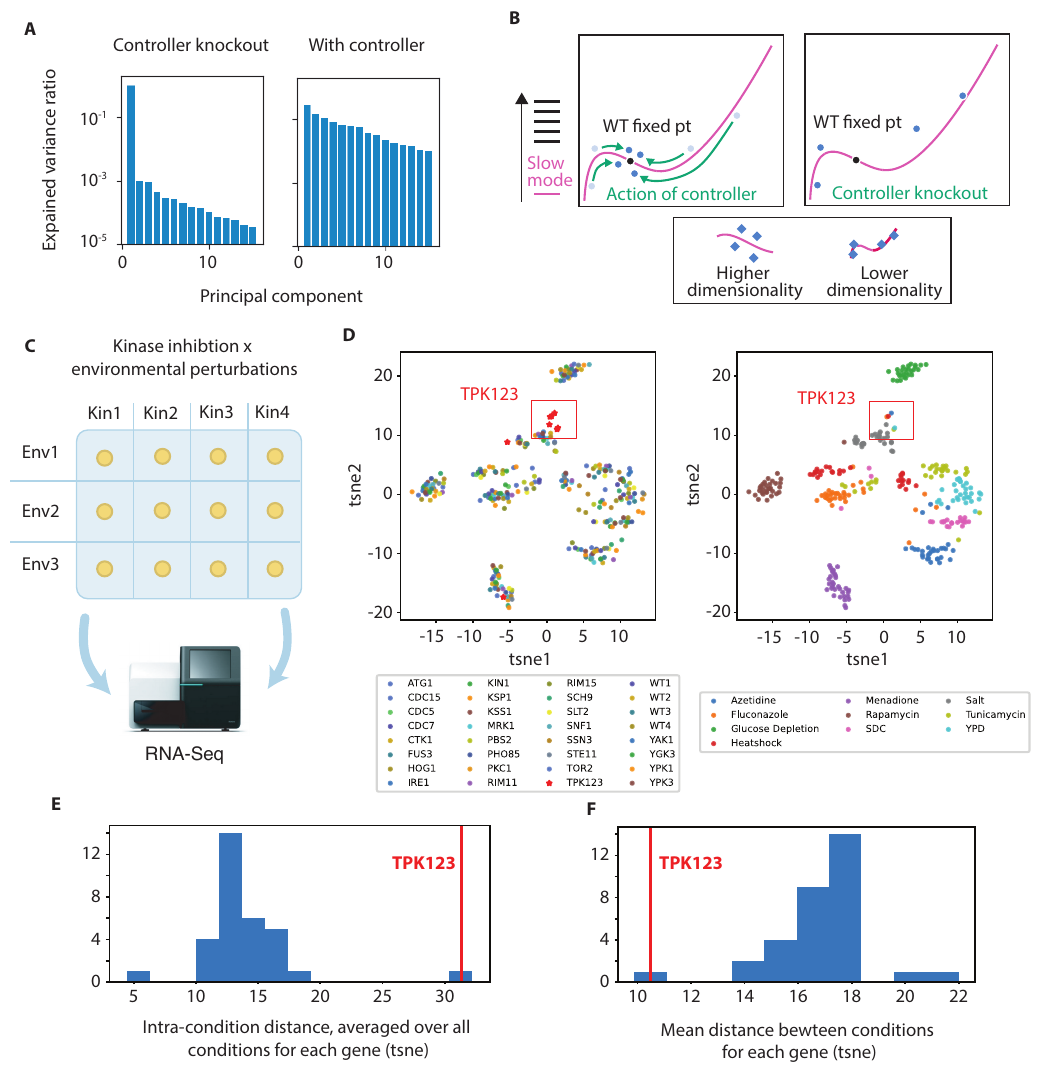} 
     \caption{Slow modes imply that knocking out a controller will, counterintuitively, reduce the dimensionality of an environmental perturbation. (a) We simulated the effects of diverse environmental perturbations on our co-evolved network and slow mode mediated simple controller with and without the controller knocked-out. We apply PCA to the simulation results, showing the reduction of dimensionality due to the knockout. (b) Without a controller, perturbations largely push the system along the slow mode, leading to a lower observed dimensionality. (c) Yeast cells subject to inhibition of 28 different kinases were grown in 10 different environmental conditions, and transcriptomic data was collected \cite{mace2020multi}. (d) Clustering of RNA-seq samples following dimensionality reduction by t-SNE plotted in two dimensions. Colors indicate different kinase inhibitions or environmental conditions, respectively. Tpk123 inhibited samples cluster, regardless of environmental condition. For other conditions, samples generally cluster on the basis of environmental perturbation, regardless of kinase knock-down. Tpk123 knock-down leads to stereotyped effect across environmental conditions; for other knockdowns, different environmental conditions have distinct effects. (e) The average distance in tsne space to all other GxE samples of the same condition was computed, and then an average of averages was computed for each genetic background and is plotted. The TPK123 value is highlighted. (f) The average distance in tsne space to all other GxE samples of the same genetic background was computed and is plotted, and the TPK123 value is highlighted. }
    \label{fig6}
\end{figure*}
\par 

 In addition to providing an explanation for the surprising effectiveness of such controllers, our model makes testable predictions about the behavior of biological systems where robustness is mediated by a slow mode. 

Our model predicts that if a controller is sensing and acting on a slow mode, even if evolution selected exclusively for environmental robustness, such a controller would facilitate mutational robustness as well.

This suggests that genes that buffer mutations tend to buffer environmental stresses as well. We analyze fitness data from high throughput yeast experiments that explore the interacting effects of mutations and environmental stresses, and show this. 

Our model also predicts that if such a controller functions by listening to and acting on a slow mode, knocking out the controller exposes a very soft mode to perturbations. Thus, in the absence of such a controller, our work predicts that the impact of perturbations should be counterintuitively lower dimensional (that is, stereotyped in the direction of that mode) than in the presence of the controller.

We analyze yeast kinase results which preliminarily add credence to this prediction. Future experiments, which quantify the dimensionality of the effect of perturbations under controller knockout, could add further support for the theory.

\begin{acknowledgments}
\end{acknowledgments}
We are grateful to Abigail Skwara, Pankaj Mehta, Mikhail Tikhonov, Terry Hwa, Shenshen Wang, Ariel Amir, Maryn Carlson, Annisa Dea, Lauren McGough, the Murugan lab, the Chan-Zuckerberg theory group for helpful discussions. This work was supported by the Chan-Zuckerberg Initiative, National Science Foundation through the Center for Living Systems (grant no. 2317138) and the NIGMS of the NIH under award number R35GM151211.

\clearpage

\bibliography{bib}

\begin{thebibliography}{42}%
\makeatletter
\providecommand \@ifxundefined [1]{%
 \@ifx{#1\undefined}
}%
\providecommand \@ifnum [1]{%
 \ifnum #1\expandafter \@firstoftwo
 \else \expandafter \@secondoftwo
 \fi
}%
\providecommand \@ifx [1]{%
 \ifx #1\expandafter \@firstoftwo
 \else \expandafter \@secondoftwo
 \fi
}%
\providecommand \natexlab [1]{#1}%
\providecommand \enquote  [1]{``#1''}%
\providecommand \bibnamefont  [1]{#1}%
\providecommand \bibfnamefont [1]{#1}%
\providecommand \citenamefont [1]{#1}%
\providecommand \href@noop [0]{\@secondoftwo}%
\providecommand \href [0]{\begingroup \@sanitize@url \@href}%
\providecommand \@href[1]{\@@startlink{#1}\@@href}%
\providecommand \@@href[1]{\endgroup#1\@@endlink}%
\providecommand \@sanitize@url [0]{\catcode `\\12\catcode `\$12\catcode `\&12\catcode `\#12\catcode `\^12\catcode `\_12\catcode `\%12\relax}%
\providecommand \@@startlink[1]{}%
\providecommand \@@endlink[0]{}%
\providecommand \url  [0]{\begingroup\@sanitize@url \@url }%
\providecommand \@url [1]{\endgroup\@href {#1}{\urlprefix }}%
\providecommand \urlprefix  [0]{URL }%
\providecommand \Eprint [0]{\href }%
\providecommand \doibase [0]{https://doi.org/}%
\providecommand \selectlanguage [0]{\@gobble}%
\providecommand \bibinfo  [0]{\@secondoftwo}%
\providecommand \bibfield  [0]{\@secondoftwo}%
\providecommand \translation [1]{[#1]}%
\providecommand \BibitemOpen [0]{}%
\providecommand \bibitemStop [0]{}%
\providecommand \bibitemNoStop [0]{.\EOS\space}%
\providecommand \EOS [0]{\spacefactor3000\relax}%
\providecommand \BibitemShut  [1]{\csname bibitem#1\endcsname}%
\let\auto@bib@innerbib\@empty
\bibitem [{\citenamefont {Milo}\ and\ \citenamefont {Phillips}(2015)}]{milo2015cell}%
  \BibitemOpen
  \bibfield  {author} {\bibinfo {author} {\bibfnamefont {R.}~\bibnamefont {Milo}}\ and\ \bibinfo {author} {\bibfnamefont {R.}~\bibnamefont {Phillips}},\ }\href@noop {} {\emph {\bibinfo {title} {Cell biology by the numbers}}}\ (\bibinfo  {publisher} {Garland Science},\ \bibinfo {year} {2015})\BibitemShut {NoStop}%
\bibitem [{\citenamefont {Wu}\ \emph {et~al.}(2022)\citenamefont {Wu}, \citenamefont {Balakrishnan}, \citenamefont {Braniff}, \citenamefont {Mori}, \citenamefont {Manzanarez}, \citenamefont {Zhang},\ and\ \citenamefont {Hwa}}]{wu2022cellular}%
  \BibitemOpen
  \bibfield  {author} {\bibinfo {author} {\bibfnamefont {C.}~\bibnamefont {Wu}}, \bibinfo {author} {\bibfnamefont {R.}~\bibnamefont {Balakrishnan}}, \bibinfo {author} {\bibfnamefont {N.}~\bibnamefont {Braniff}}, \bibinfo {author} {\bibfnamefont {M.}~\bibnamefont {Mori}}, \bibinfo {author} {\bibfnamefont {G.}~\bibnamefont {Manzanarez}}, \bibinfo {author} {\bibfnamefont {Z.}~\bibnamefont {Zhang}},\ and\ \bibinfo {author} {\bibfnamefont {T.}~\bibnamefont {Hwa}},\ }\bibfield  {title} {\bibinfo {title} {Cellular perception of growth rate and the mechanistic origin of bacterial growth law},\ }\href@noop {} {\bibfield  {journal} {\bibinfo  {journal} {Proceedings of the National Academy of Sciences}\ }\textbf {\bibinfo {volume} {119}},\ \bibinfo {pages} {e2201585119} (\bibinfo {year} {2022})}\BibitemShut {NoStop}%
\bibitem [{\citenamefont {Traxler}\ \emph {et~al.}(2008)\citenamefont {Traxler}, \citenamefont {Summers}, \citenamefont {Nguyen}, \citenamefont {Zacharia}, \citenamefont {Hightower}, \citenamefont {Smith},\ and\ \citenamefont {Conway}}]{traxler2008global}%
  \BibitemOpen
  \bibfield  {author} {\bibinfo {author} {\bibfnamefont {M.~F.}\ \bibnamefont {Traxler}}, \bibinfo {author} {\bibfnamefont {S.~M.}\ \bibnamefont {Summers}}, \bibinfo {author} {\bibfnamefont {H.-T.}\ \bibnamefont {Nguyen}}, \bibinfo {author} {\bibfnamefont {V.~M.}\ \bibnamefont {Zacharia}}, \bibinfo {author} {\bibfnamefont {G.~A.}\ \bibnamefont {Hightower}}, \bibinfo {author} {\bibfnamefont {J.~T.}\ \bibnamefont {Smith}},\ and\ \bibinfo {author} {\bibfnamefont {T.}~\bibnamefont {Conway}},\ }\bibfield  {title} {\bibinfo {title} {The global, ppgpp-mediated stringent response to amino acid starvation in escherichia coli},\ }\href@noop {} {\bibfield  {journal} {\bibinfo  {journal} {Molecular microbiology}\ }\textbf {\bibinfo {volume} {68}},\ \bibinfo {pages} {1128} (\bibinfo {year} {2008})}\BibitemShut {NoStop}%
\bibitem [{\citenamefont {Magnusson}\ \emph {et~al.}(2005)\citenamefont {Magnusson}, \citenamefont {Farewell},\ and\ \citenamefont {Nystr{\"o}m}}]{magnusson2005ppgpp}%
  \BibitemOpen
  \bibfield  {author} {\bibinfo {author} {\bibfnamefont {L.~U.}\ \bibnamefont {Magnusson}}, \bibinfo {author} {\bibfnamefont {A.}~\bibnamefont {Farewell}},\ and\ \bibinfo {author} {\bibfnamefont {T.}~\bibnamefont {Nystr{\"o}m}},\ }\bibfield  {title} {\bibinfo {title} {ppgpp: a global regulator in escherichia coli},\ }\href@noop {} {\bibfield  {journal} {\bibinfo  {journal} {Trends in microbiology}\ }\textbf {\bibinfo {volume} {13}},\ \bibinfo {pages} {236} (\bibinfo {year} {2005})}\BibitemShut {NoStop}%
\bibitem [{\citenamefont {Conrad}\ \emph {et~al.}(2014)\citenamefont {Conrad}, \citenamefont {Schothorst}, \citenamefont {Kankipati}, \citenamefont {Van~Zeebroeck}, \citenamefont {Rubio-Texeira},\ and\ \citenamefont {Thevelein}}]{conrad2014nutrient}%
  \BibitemOpen
  \bibfield  {author} {\bibinfo {author} {\bibfnamefont {M.}~\bibnamefont {Conrad}}, \bibinfo {author} {\bibfnamefont {J.}~\bibnamefont {Schothorst}}, \bibinfo {author} {\bibfnamefont {H.~N.}\ \bibnamefont {Kankipati}}, \bibinfo {author} {\bibfnamefont {G.}~\bibnamefont {Van~Zeebroeck}}, \bibinfo {author} {\bibfnamefont {M.}~\bibnamefont {Rubio-Texeira}},\ and\ \bibinfo {author} {\bibfnamefont {J.~M.}\ \bibnamefont {Thevelein}},\ }\bibfield  {title} {\bibinfo {title} {Nutrient sensing and signaling in the yeast saccharomyces cerevisiae},\ }\href@noop {} {\bibfield  {journal} {\bibinfo  {journal} {FEMS microbiology reviews}\ }\textbf {\bibinfo {volume} {38}},\ \bibinfo {pages} {254} (\bibinfo {year} {2014})}\BibitemShut {NoStop}%
\bibitem [{\citenamefont {Thevelein}\ and\ \citenamefont {De~Winde}(1999)}]{thevelein1999novel}%
  \BibitemOpen
  \bibfield  {author} {\bibinfo {author} {\bibfnamefont {J.~M.}\ \bibnamefont {Thevelein}}\ and\ \bibinfo {author} {\bibfnamefont {J.~H.}\ \bibnamefont {De~Winde}},\ }\bibfield  {title} {\bibinfo {title} {Novel sensing mechanisms and targets for the camp--protein kinase a pathway in the yeast saccharomyces cerevisiae},\ }\href@noop {} {\bibfield  {journal} {\bibinfo  {journal} {Molecular microbiology}\ }\textbf {\bibinfo {volume} {33}},\ \bibinfo {pages} {904} (\bibinfo {year} {1999})}\BibitemShut {NoStop}%
\bibitem [{\citenamefont {Gottesman}(2019)}]{gottesman2019trouble}%
  \BibitemOpen
  \bibfield  {author} {\bibinfo {author} {\bibfnamefont {S.}~\bibnamefont {Gottesman}},\ }\bibfield  {title} {\bibinfo {title} {Trouble is coming: Signaling pathways that regulate general stress responses in bacteria},\ }\href@noop {} {\bibfield  {journal} {\bibinfo  {journal} {Journal of Biological Chemistry}\ }\textbf {\bibinfo {volume} {294}},\ \bibinfo {pages} {11685} (\bibinfo {year} {2019})}\BibitemShut {NoStop}%
\bibitem [{\citenamefont {Serezani}\ \emph {et~al.}(2008)\citenamefont {Serezani}, \citenamefont {Ballinger}, \citenamefont {Aronoff},\ and\ \citenamefont {Peters-Golden}}]{serezani2008cyclic}%
  \BibitemOpen
  \bibfield  {author} {\bibinfo {author} {\bibfnamefont {C.~H.}\ \bibnamefont {Serezani}}, \bibinfo {author} {\bibfnamefont {M.~N.}\ \bibnamefont {Ballinger}}, \bibinfo {author} {\bibfnamefont {D.~M.}\ \bibnamefont {Aronoff}},\ and\ \bibinfo {author} {\bibfnamefont {M.}~\bibnamefont {Peters-Golden}},\ }\bibfield  {title} {\bibinfo {title} {Cyclic amp: master regulator of innate immune cell function},\ }\href@noop {} {\bibfield  {journal} {\bibinfo  {journal} {American journal of respiratory cell and molecular biology}\ }\textbf {\bibinfo {volume} {39}},\ \bibinfo {pages} {127} (\bibinfo {year} {2008})}\BibitemShut {NoStop}%
\bibitem [{\citenamefont {Bahar}(2010)}]{bahar2010functional}%
  \BibitemOpen
  \bibfield  {author} {\bibinfo {author} {\bibfnamefont {I.}~\bibnamefont {Bahar}},\ }\bibfield  {title} {\bibinfo {title} {On the functional significance of soft modes predicted by coarse-grained models for membrane proteins},\ }\href@noop {} {\bibfield  {journal} {\bibinfo  {journal} {Journal of General Physiology}\ }\textbf {\bibinfo {volume} {135}},\ \bibinfo {pages} {563} (\bibinfo {year} {2010})}\BibitemShut {NoStop}%
\bibitem [{\citenamefont {Leo-Macias}\ \emph {et~al.}(2005)\citenamefont {Leo-Macias}, \citenamefont {Lopez-Romero}, \citenamefont {Lupyan}, \citenamefont {Zerbino},\ and\ \citenamefont {Ortiz}}]{leo2005analysis}%
  \BibitemOpen
  \bibfield  {author} {\bibinfo {author} {\bibfnamefont {A.}~\bibnamefont {Leo-Macias}}, \bibinfo {author} {\bibfnamefont {P.}~\bibnamefont {Lopez-Romero}}, \bibinfo {author} {\bibfnamefont {D.}~\bibnamefont {Lupyan}}, \bibinfo {author} {\bibfnamefont {D.}~\bibnamefont {Zerbino}},\ and\ \bibinfo {author} {\bibfnamefont {A.~R.}\ \bibnamefont {Ortiz}},\ }\bibfield  {title} {\bibinfo {title} {An analysis of core deformations in protein superfamilies},\ }\href@noop {} {\bibfield  {journal} {\bibinfo  {journal} {Biophysical journal}\ }\textbf {\bibinfo {volume} {88}},\ \bibinfo {pages} {1291} (\bibinfo {year} {2005})}\BibitemShut {NoStop}%
\bibitem [{\citenamefont {Russo}\ \emph {et~al.}(2024)\citenamefont {Russo}, \citenamefont {Husain},\ and\ \citenamefont {Murugan}}]{russo2024soft}%
  \BibitemOpen
  \bibfield  {author} {\bibinfo {author} {\bibfnamefont {C.~J.}\ \bibnamefont {Russo}}, \bibinfo {author} {\bibfnamefont {K.}~\bibnamefont {Husain}},\ and\ \bibinfo {author} {\bibfnamefont {A.}~\bibnamefont {Murugan}},\ }\bibfield  {title} {\bibinfo {title} {Soft modes as a predictive framework for low-dimensional biological systems across scales},\ }\href@noop {} {\bibfield  {journal} {\bibinfo  {journal} {Annual Review of Biophysics}\ }\textbf {\bibinfo {volume} {54}} (\bibinfo {year} {2024})}\BibitemShut {NoStop}%
\bibitem [{\citenamefont {Husain}\ and\ \citenamefont {Murugan}(2020)}]{husain2020physical}%
  \BibitemOpen
  \bibfield  {author} {\bibinfo {author} {\bibfnamefont {K.}~\bibnamefont {Husain}}\ and\ \bibinfo {author} {\bibfnamefont {A.}~\bibnamefont {Murugan}},\ }\bibfield  {title} {\bibinfo {title} {Physical constraints on epistasis},\ }\href@noop {} {\bibfield  {journal} {\bibinfo  {journal} {Molecular Biology and Evolution}\ }\textbf {\bibinfo {volume} {37}},\ \bibinfo {pages} {2865} (\bibinfo {year} {2020})}\BibitemShut {NoStop}%
\bibitem [{\citenamefont {Costanzo}\ \emph {et~al.}(2016)\citenamefont {Costanzo}, \citenamefont {VanderSluis}, \citenamefont {Koch}, \citenamefont {Baryshnikova}, \citenamefont {Pons}, \citenamefont {Tan}, \citenamefont {Wang}, \citenamefont {Usaj}, \citenamefont {Hanchard}, \citenamefont {Lee} \emph {et~al.}}]{costanzo2016global}%
  \BibitemOpen
  \bibfield  {author} {\bibinfo {author} {\bibfnamefont {M.}~\bibnamefont {Costanzo}}, \bibinfo {author} {\bibfnamefont {B.}~\bibnamefont {VanderSluis}}, \bibinfo {author} {\bibfnamefont {E.~N.}\ \bibnamefont {Koch}}, \bibinfo {author} {\bibfnamefont {A.}~\bibnamefont {Baryshnikova}}, \bibinfo {author} {\bibfnamefont {C.}~\bibnamefont {Pons}}, \bibinfo {author} {\bibfnamefont {G.}~\bibnamefont {Tan}}, \bibinfo {author} {\bibfnamefont {W.}~\bibnamefont {Wang}}, \bibinfo {author} {\bibfnamefont {M.}~\bibnamefont {Usaj}}, \bibinfo {author} {\bibfnamefont {J.}~\bibnamefont {Hanchard}}, \bibinfo {author} {\bibfnamefont {S.~D.}\ \bibnamefont {Lee}}, \emph {et~al.},\ }\bibfield  {title} {\bibinfo {title} {A global genetic interaction network maps a wiring diagram of cellular function},\ }\href@noop {} {\bibfield  {journal} {\bibinfo  {journal} {Science}\ }\textbf {\bibinfo {volume} {353}},\ \bibinfo {pages} {aaf1420} (\bibinfo {year} {2016})}\BibitemShut {NoStop}%
\bibitem [{\citenamefont {Costanzo}\ \emph {et~al.}(2021)\citenamefont {Costanzo}, \citenamefont {Hou}, \citenamefont {Messier}, \citenamefont {Nelson}, \citenamefont {Rahman}, \citenamefont {VanderSluis}, \citenamefont {Wang}, \citenamefont {Pons}, \citenamefont {Ross}, \citenamefont {U{\v{s}}aj} \emph {et~al.}}]{costanzo2021environmental}%
  \BibitemOpen
  \bibfield  {author} {\bibinfo {author} {\bibfnamefont {M.}~\bibnamefont {Costanzo}}, \bibinfo {author} {\bibfnamefont {J.}~\bibnamefont {Hou}}, \bibinfo {author} {\bibfnamefont {V.}~\bibnamefont {Messier}}, \bibinfo {author} {\bibfnamefont {J.}~\bibnamefont {Nelson}}, \bibinfo {author} {\bibfnamefont {M.}~\bibnamefont {Rahman}}, \bibinfo {author} {\bibfnamefont {B.}~\bibnamefont {VanderSluis}}, \bibinfo {author} {\bibfnamefont {W.}~\bibnamefont {Wang}}, \bibinfo {author} {\bibfnamefont {C.}~\bibnamefont {Pons}}, \bibinfo {author} {\bibfnamefont {C.}~\bibnamefont {Ross}}, \bibinfo {author} {\bibfnamefont {M.}~\bibnamefont {U{\v{s}}aj}}, \emph {et~al.},\ }\bibfield  {title} {\bibinfo {title} {Environmental robustness of the global yeast genetic interaction network},\ }\href@noop {} {\bibfield  {journal} {\bibinfo  {journal} {Science}\ }\textbf {\bibinfo {volume} {372}},\ \bibinfo {pages} {eabf8424} (\bibinfo {year} {2021})}\BibitemShut {NoStop}%
\bibitem [{\citenamefont {Mitchell}\ \emph {et~al.}(2016)\citenamefont {Mitchell}, \citenamefont {Tlusty},\ and\ \citenamefont {Leibler}}]{mitchell2016strain}%
  \BibitemOpen
  \bibfield  {author} {\bibinfo {author} {\bibfnamefont {M.~R.}\ \bibnamefont {Mitchell}}, \bibinfo {author} {\bibfnamefont {T.}~\bibnamefont {Tlusty}},\ and\ \bibinfo {author} {\bibfnamefont {S.}~\bibnamefont {Leibler}},\ }\bibfield  {title} {\bibinfo {title} {Strain analysis of protein structures and low dimensionality of mechanical allosteric couplings},\ }\href@noop {} {\bibfield  {journal} {\bibinfo  {journal} {Proceedings of the National Academy of Sciences}\ }\textbf {\bibinfo {volume} {113}},\ \bibinfo {pages} {E5847} (\bibinfo {year} {2016})}\BibitemShut {NoStop}%
\bibitem [{\citenamefont {Rocks}\ \emph {et~al.}(2017)\citenamefont {Rocks}, \citenamefont {Pashine}, \citenamefont {Bischofberger}, \citenamefont {Goodrich}, \citenamefont {Liu},\ and\ \citenamefont {Nagel}}]{rocks2017designing}%
  \BibitemOpen
  \bibfield  {author} {\bibinfo {author} {\bibfnamefont {J.~W.}\ \bibnamefont {Rocks}}, \bibinfo {author} {\bibfnamefont {N.}~\bibnamefont {Pashine}}, \bibinfo {author} {\bibfnamefont {I.}~\bibnamefont {Bischofberger}}, \bibinfo {author} {\bibfnamefont {C.~P.}\ \bibnamefont {Goodrich}}, \bibinfo {author} {\bibfnamefont {A.~J.}\ \bibnamefont {Liu}},\ and\ \bibinfo {author} {\bibfnamefont {S.~R.}\ \bibnamefont {Nagel}},\ }\bibfield  {title} {\bibinfo {title} {Designing allostery-inspired response in mechanical networks},\ }\href@noop {} {\bibfield  {journal} {\bibinfo  {journal} {Proceedings of the National Academy of Sciences}\ }\textbf {\bibinfo {volume} {114}},\ \bibinfo {pages} {2520} (\bibinfo {year} {2017})}\BibitemShut {NoStop}%
\bibitem [{\citenamefont {Campitelli}\ \emph {et~al.}(2020)\citenamefont {Campitelli}, \citenamefont {Modi}, \citenamefont {Kumar},\ and\ \citenamefont {Ozkan}}]{campitelli2020role}%
  \BibitemOpen
  \bibfield  {author} {\bibinfo {author} {\bibfnamefont {P.}~\bibnamefont {Campitelli}}, \bibinfo {author} {\bibfnamefont {T.}~\bibnamefont {Modi}}, \bibinfo {author} {\bibfnamefont {S.}~\bibnamefont {Kumar}},\ and\ \bibinfo {author} {\bibfnamefont {S.~B.}\ \bibnamefont {Ozkan}},\ }\bibfield  {title} {\bibinfo {title} {The role of conformational dynamics and allostery in modulating protein evolution},\ }\href@noop {} {\bibfield  {journal} {\bibinfo  {journal} {Annual review of biophysics}\ }\textbf {\bibinfo {volume} {49}},\ \bibinfo {pages} {267} (\bibinfo {year} {2020})}\BibitemShut {NoStop}%
\bibitem [{\citenamefont {Ravasio}\ \emph {et~al.}(2019)\citenamefont {Ravasio}, \citenamefont {Flatt}, \citenamefont {Yan}, \citenamefont {Zamuner}, \citenamefont {Brito},\ and\ \citenamefont {Wyart}}]{ravasio2019mechanics}%
  \BibitemOpen
  \bibfield  {author} {\bibinfo {author} {\bibfnamefont {R.}~\bibnamefont {Ravasio}}, \bibinfo {author} {\bibfnamefont {S.~M.}\ \bibnamefont {Flatt}}, \bibinfo {author} {\bibfnamefont {L.}~\bibnamefont {Yan}}, \bibinfo {author} {\bibfnamefont {S.}~\bibnamefont {Zamuner}}, \bibinfo {author} {\bibfnamefont {C.}~\bibnamefont {Brito}},\ and\ \bibinfo {author} {\bibfnamefont {M.}~\bibnamefont {Wyart}},\ }\bibfield  {title} {\bibinfo {title} {Mechanics of allostery: contrasting the induced fit and population shift scenarios},\ }\href@noop {} {\bibfield  {journal} {\bibinfo  {journal} {Biophysical journal}\ }\textbf {\bibinfo {volume} {117}},\ \bibinfo {pages} {1954} (\bibinfo {year} {2019})}\BibitemShut {NoStop}%
\bibitem [{\citenamefont {Raman}\ \emph {et~al.}(2016)\citenamefont {Raman}, \citenamefont {White},\ and\ \citenamefont {Ranganathan}}]{raman2016origins}%
  \BibitemOpen
  \bibfield  {author} {\bibinfo {author} {\bibfnamefont {A.~S.}\ \bibnamefont {Raman}}, \bibinfo {author} {\bibfnamefont {K.~I.}\ \bibnamefont {White}},\ and\ \bibinfo {author} {\bibfnamefont {R.}~\bibnamefont {Ranganathan}},\ }\bibfield  {title} {\bibinfo {title} {Origins of allostery and evolvability in proteins: a case study},\ }\href@noop {} {\bibfield  {journal} {\bibinfo  {journal} {Cell}\ }\textbf {\bibinfo {volume} {166}},\ \bibinfo {pages} {468} (\bibinfo {year} {2016})}\BibitemShut {NoStop}%
\bibitem [{\citenamefont {Friedlander}\ \emph {et~al.}(2015)\citenamefont {Friedlander}, \citenamefont {Mayo}, \citenamefont {Tlusty},\ and\ \citenamefont {Alon}}]{friedlander2015evolution}%
  \BibitemOpen
  \bibfield  {author} {\bibinfo {author} {\bibfnamefont {T.}~\bibnamefont {Friedlander}}, \bibinfo {author} {\bibfnamefont {A.~E.}\ \bibnamefont {Mayo}}, \bibinfo {author} {\bibfnamefont {T.}~\bibnamefont {Tlusty}},\ and\ \bibinfo {author} {\bibfnamefont {U.}~\bibnamefont {Alon}},\ }\bibfield  {title} {\bibinfo {title} {Evolution of bow-tie architectures in biology},\ }\href@noop {} {\bibfield  {journal} {\bibinfo  {journal} {PLoS computational biology}\ }\textbf {\bibinfo {volume} {11}},\ \bibinfo {pages} {e1004055} (\bibinfo {year} {2015})}\BibitemShut {NoStop}%
\bibitem [{\citenamefont {Furusawa}\ and\ \citenamefont {Kaneko}(2018)}]{furusawa2018formation}%
  \BibitemOpen
  \bibfield  {author} {\bibinfo {author} {\bibfnamefont {C.}~\bibnamefont {Furusawa}}\ and\ \bibinfo {author} {\bibfnamefont {K.}~\bibnamefont {Kaneko}},\ }\bibfield  {title} {\bibinfo {title} {Formation of dominant mode by evolution in biological systems},\ }\href@noop {} {\bibfield  {journal} {\bibinfo  {journal} {Physical Review E}\ }\textbf {\bibinfo {volume} {97}},\ \bibinfo {pages} {042410} (\bibinfo {year} {2018})}\BibitemShut {NoStop}%
\bibitem [{\citenamefont {Yi}\ \emph {et~al.}(2000)\citenamefont {Yi}, \citenamefont {Huang}, \citenamefont {Simon},\ and\ \citenamefont {Doyle}}]{yi2000robust}%
  \BibitemOpen
  \bibfield  {author} {\bibinfo {author} {\bibfnamefont {T.-M.}\ \bibnamefont {Yi}}, \bibinfo {author} {\bibfnamefont {Y.}~\bibnamefont {Huang}}, \bibinfo {author} {\bibfnamefont {M.~I.}\ \bibnamefont {Simon}},\ and\ \bibinfo {author} {\bibfnamefont {J.}~\bibnamefont {Doyle}},\ }\bibfield  {title} {\bibinfo {title} {Robust perfect adaptation in bacterial chemotaxis through integral feedback control},\ }\href@noop {} {\bibfield  {journal} {\bibinfo  {journal} {Proceedings of the National Academy of Sciences}\ }\textbf {\bibinfo {volume} {97}},\ \bibinfo {pages} {4649} (\bibinfo {year} {2000})}\BibitemShut {NoStop}%
\bibitem [{\citenamefont {Somvanshi}\ \emph {et~al.}(2015)\citenamefont {Somvanshi}, \citenamefont {Patel}, \citenamefont {Bhartiya},\ and\ \citenamefont {Venkatesh}}]{somvanshi2015implementation}%
  \BibitemOpen
  \bibfield  {author} {\bibinfo {author} {\bibfnamefont {P.~R.}\ \bibnamefont {Somvanshi}}, \bibinfo {author} {\bibfnamefont {A.~K.}\ \bibnamefont {Patel}}, \bibinfo {author} {\bibfnamefont {S.}~\bibnamefont {Bhartiya}},\ and\ \bibinfo {author} {\bibfnamefont {K.}~\bibnamefont {Venkatesh}},\ }\bibfield  {title} {\bibinfo {title} {Implementation of integral feedback control in biological systems},\ }\href@noop {} {\bibfield  {journal} {\bibinfo  {journal} {Wiley Interdisciplinary Reviews: Systems Biology and Medicine}\ }\textbf {\bibinfo {volume} {7}},\ \bibinfo {pages} {301} (\bibinfo {year} {2015})}\BibitemShut {NoStop}%
\bibitem [{\citenamefont {Roy}\ and\ \citenamefont {Vetterli}(2007)}]{roy2007effective}%
  \BibitemOpen
  \bibfield  {author} {\bibinfo {author} {\bibfnamefont {O.}~\bibnamefont {Roy}}\ and\ \bibinfo {author} {\bibfnamefont {M.}~\bibnamefont {Vetterli}},\ }\bibfield  {title} {\bibinfo {title} {The effective rank: A measure of effective dimensionality},\ }in\ \href@noop {} {\emph {\bibinfo {booktitle} {2007 15th European signal processing conference}}}\ (\bibinfo {organization} {IEEE},\ \bibinfo {year} {2007})\ pp.\ \bibinfo {pages} {606--610}\BibitemShut {NoStop}%
\bibitem [{\citenamefont {Rutherford}\ and\ \citenamefont {Lindquist}(1998)}]{rutherford1998hsp90}%
  \BibitemOpen
  \bibfield  {author} {\bibinfo {author} {\bibfnamefont {S.~L.}\ \bibnamefont {Rutherford}}\ and\ \bibinfo {author} {\bibfnamefont {S.}~\bibnamefont {Lindquist}},\ }\bibfield  {title} {\bibinfo {title} {Hsp90 as a capacitor for morphological evolution},\ }\href@noop {} {\bibfield  {journal} {\bibinfo  {journal} {Nature}\ }\textbf {\bibinfo {volume} {396}},\ \bibinfo {pages} {336} (\bibinfo {year} {1998})}\BibitemShut {NoStop}%
\bibitem [{\citenamefont {Queitsch}\ \emph {et~al.}(2002)\citenamefont {Queitsch}, \citenamefont {Sangster},\ and\ \citenamefont {Lindquist}}]{queitsch2002hsp90}%
  \BibitemOpen
  \bibfield  {author} {\bibinfo {author} {\bibfnamefont {C.}~\bibnamefont {Queitsch}}, \bibinfo {author} {\bibfnamefont {T.~A.}\ \bibnamefont {Sangster}},\ and\ \bibinfo {author} {\bibfnamefont {S.}~\bibnamefont {Lindquist}},\ }\bibfield  {title} {\bibinfo {title} {Hsp90 as a capacitor of phenotypic variation},\ }\href@noop {} {\bibfield  {journal} {\bibinfo  {journal} {Nature}\ }\textbf {\bibinfo {volume} {417}},\ \bibinfo {pages} {618} (\bibinfo {year} {2002})}\BibitemShut {NoStop}%
\bibitem [{\citenamefont {Yeyati}\ and\ \citenamefont {van Heyningen}(2008)}]{yeyati2008incapacitating}%
  \BibitemOpen
  \bibfield  {author} {\bibinfo {author} {\bibfnamefont {P.~L.}\ \bibnamefont {Yeyati}}\ and\ \bibinfo {author} {\bibfnamefont {V.}~\bibnamefont {van Heyningen}},\ }\bibfield  {title} {\bibinfo {title} {Incapacitating the evolutionary capacitor: Hsp90 modulation of disease},\ }\href@noop {} {\bibfield  {journal} {\bibinfo  {journal} {Current opinion in genetics \& development}\ }\textbf {\bibinfo {volume} {18}},\ \bibinfo {pages} {264} (\bibinfo {year} {2008})}\BibitemShut {NoStop}%
\bibitem [{\citenamefont {Costanzo}\ \emph {et~al.}(2010)\citenamefont {Costanzo}, \citenamefont {Baryshnikova}, \citenamefont {Bellay}, \citenamefont {Kim}, \citenamefont {Spear}, \citenamefont {Sevier}, \citenamefont {Ding}, \citenamefont {Koh}, \citenamefont {Toufighi}, \citenamefont {Mostafavi} \emph {et~al.}}]{costanzo2010genetic}%
  \BibitemOpen
  \bibfield  {author} {\bibinfo {author} {\bibfnamefont {M.}~\bibnamefont {Costanzo}}, \bibinfo {author} {\bibfnamefont {A.}~\bibnamefont {Baryshnikova}}, \bibinfo {author} {\bibfnamefont {J.}~\bibnamefont {Bellay}}, \bibinfo {author} {\bibfnamefont {Y.}~\bibnamefont {Kim}}, \bibinfo {author} {\bibfnamefont {E.~D.}\ \bibnamefont {Spear}}, \bibinfo {author} {\bibfnamefont {C.~S.}\ \bibnamefont {Sevier}}, \bibinfo {author} {\bibfnamefont {H.}~\bibnamefont {Ding}}, \bibinfo {author} {\bibfnamefont {J.~L.}\ \bibnamefont {Koh}}, \bibinfo {author} {\bibfnamefont {K.}~\bibnamefont {Toufighi}}, \bibinfo {author} {\bibfnamefont {S.}~\bibnamefont {Mostafavi}}, \emph {et~al.},\ }\bibfield  {title} {\bibinfo {title} {The genetic landscape of a cell},\ }\href@noop {} {\bibfield  {journal} {\bibinfo  {journal} {science}\ }\textbf {\bibinfo {volume} {327}},\ \bibinfo {pages} {425} (\bibinfo {year} {2010})}\BibitemShut {NoStop}%
\bibitem [{\citenamefont {Ballou}\ and\ \citenamefont {Lin}(2008)}]{ballou2008rapamycin}%
  \BibitemOpen
  \bibfield  {author} {\bibinfo {author} {\bibfnamefont {L.~M.}\ \bibnamefont {Ballou}}\ and\ \bibinfo {author} {\bibfnamefont {R.~Z.}\ \bibnamefont {Lin}},\ }\bibfield  {title} {\bibinfo {title} {Rapamycin and mtor kinase inhibitors},\ }\href@noop {} {\bibfield  {journal} {\bibinfo  {journal} {Journal of chemical biology}\ }\textbf {\bibinfo {volume} {1}},\ \bibinfo {pages} {27} (\bibinfo {year} {2008})}\BibitemShut {NoStop}%
\bibitem [{\citenamefont {Mace}\ \emph {et~al.}(2020)\citenamefont {Mace}, \citenamefont {Krakowiak}, \citenamefont {El-Samad},\ and\ \citenamefont {Pincus}}]{mace2020multi}%
  \BibitemOpen
  \bibfield  {author} {\bibinfo {author} {\bibfnamefont {K.}~\bibnamefont {Mace}}, \bibinfo {author} {\bibfnamefont {J.}~\bibnamefont {Krakowiak}}, \bibinfo {author} {\bibfnamefont {H.}~\bibnamefont {El-Samad}},\ and\ \bibinfo {author} {\bibfnamefont {D.}~\bibnamefont {Pincus}},\ }\bibfield  {title} {\bibinfo {title} {Multi-kinase control of environmental stress responsive transcription},\ }\href@noop {} {\bibfield  {journal} {\bibinfo  {journal} {PLoS One}\ }\textbf {\bibinfo {volume} {15}},\ \bibinfo {pages} {e0230246} (\bibinfo {year} {2020})}\BibitemShut {NoStop}%
\bibitem [{\citenamefont {Love}\ \emph {et~al.}(2014)\citenamefont {Love}, \citenamefont {Anders},\ and\ \citenamefont {Huber}}]{love2014differential}%
  \BibitemOpen
  \bibfield  {author} {\bibinfo {author} {\bibfnamefont {M.}~\bibnamefont {Love}}, \bibinfo {author} {\bibfnamefont {S.}~\bibnamefont {Anders}},\ and\ \bibinfo {author} {\bibfnamefont {W.}~\bibnamefont {Huber}},\ }\bibfield  {title} {\bibinfo {title} {Differential analysis of count data--the deseq2 package},\ }\href@noop {} {\bibfield  {journal} {\bibinfo  {journal} {Genome Biol}\ }\textbf {\bibinfo {volume} {15}},\ \bibinfo {pages} {10} (\bibinfo {year} {2014})}\BibitemShut {NoStop}%
\bibitem [{\citenamefont {Van~der Maaten}\ and\ \citenamefont {Hinton}(2008)}]{van2008visualizing}%
  \BibitemOpen
  \bibfield  {author} {\bibinfo {author} {\bibfnamefont {L.}~\bibnamefont {Van~der Maaten}}\ and\ \bibinfo {author} {\bibfnamefont {G.}~\bibnamefont {Hinton}},\ }\bibfield  {title} {\bibinfo {title} {Visualizing data using t-sne.},\ }\href@noop {} {\bibfield  {journal} {\bibinfo  {journal} {Journal of machine learning research}\ }\textbf {\bibinfo {volume} {9}} (\bibinfo {year} {2008})}\BibitemShut {NoStop}%
\bibitem [{\citenamefont {Eckmann}\ and\ \citenamefont {Tlusty}(2021)}]{eckmann2021dimensional}%
  \BibitemOpen
  \bibfield  {author} {\bibinfo {author} {\bibfnamefont {J.-P.}\ \bibnamefont {Eckmann}}\ and\ \bibinfo {author} {\bibfnamefont {T.}~\bibnamefont {Tlusty}},\ }\bibfield  {title} {\bibinfo {title} {Dimensional reduction in complex living systems: Where, why, and how},\ }\href@noop {} {\bibfield  {journal} {\bibinfo  {journal} {BioEssays}\ }\textbf {\bibinfo {volume} {43}},\ \bibinfo {pages} {2100062} (\bibinfo {year} {2021})}\BibitemShut {NoStop}%
\bibitem [{\citenamefont {Raimondi}\ \emph {et~al.}(2010)\citenamefont {Raimondi}, \citenamefont {Orozco},\ and\ \citenamefont {Fanelli}}]{Raimondi2010}%
  \BibitemOpen
  \bibfield  {author} {\bibinfo {author} {\bibfnamefont {F.}~\bibnamefont {Raimondi}}, \bibinfo {author} {\bibfnamefont {M.}~\bibnamefont {Orozco}},\ and\ \bibinfo {author} {\bibfnamefont {F.}~\bibnamefont {Fanelli}},\ }\bibfield  {title} {\bibinfo {title} {{Deciphering the Deformation Modes Associated with Function Retention and Specialization in Members of the Ras Superfamily}},\ }\href {https://doi.org/10.1016/j.str.2009.12.015} {\bibfield  {journal} {\bibinfo  {journal} {Structure}\ }\textbf {\bibinfo {volume} {18}},\ \bibinfo {pages} {402} (\bibinfo {year} {2010})}\BibitemShut {NoStop}%
\bibitem [{\citenamefont {Sastry}\ \emph {et~al.}(2019)\citenamefont {Sastry}, \citenamefont {Gao}, \citenamefont {Szubin}, \citenamefont {Hefner}, \citenamefont {Xu}, \citenamefont {Kim}, \citenamefont {Choudhary}, \citenamefont {Yang}, \citenamefont {King},\ and\ \citenamefont {Palsson}}]{sastry2019escherichia}%
  \BibitemOpen
  \bibfield  {author} {\bibinfo {author} {\bibfnamefont {A.~V.}\ \bibnamefont {Sastry}}, \bibinfo {author} {\bibfnamefont {Y.}~\bibnamefont {Gao}}, \bibinfo {author} {\bibfnamefont {R.}~\bibnamefont {Szubin}}, \bibinfo {author} {\bibfnamefont {Y.}~\bibnamefont {Hefner}}, \bibinfo {author} {\bibfnamefont {S.}~\bibnamefont {Xu}}, \bibinfo {author} {\bibfnamefont {D.}~\bibnamefont {Kim}}, \bibinfo {author} {\bibfnamefont {K.~S.}\ \bibnamefont {Choudhary}}, \bibinfo {author} {\bibfnamefont {L.}~\bibnamefont {Yang}}, \bibinfo {author} {\bibfnamefont {Z.~A.}\ \bibnamefont {King}},\ and\ \bibinfo {author} {\bibfnamefont {B.~O.}\ \bibnamefont {Palsson}},\ }\bibfield  {title} {\bibinfo {title} {The escherichia coli transcriptome mostly consists of independently regulated modules},\ }\href@noop {} {\bibfield  {journal} {\bibinfo  {journal} {Nature communications}\ }\textbf {\bibinfo {volume} {10}},\ \bibinfo {pages} {5536} (\bibinfo {year} {2019})}\BibitemShut {NoStop}%
\bibitem [{\citenamefont {Schmidt}\ \emph {et~al.}(2016)\citenamefont {Schmidt}, \citenamefont {Kochanowski}, \citenamefont {Vedelaar}, \citenamefont {Ahrn{\'e}}, \citenamefont {Volkmer}, \citenamefont {Callipo}, \citenamefont {Knoops}, \citenamefont {Bauer}, \citenamefont {Aebersold},\ and\ \citenamefont {Heinemann}}]{schmidt2016quantitative}%
  \BibitemOpen
  \bibfield  {author} {\bibinfo {author} {\bibfnamefont {A.}~\bibnamefont {Schmidt}}, \bibinfo {author} {\bibfnamefont {K.}~\bibnamefont {Kochanowski}}, \bibinfo {author} {\bibfnamefont {S.}~\bibnamefont {Vedelaar}}, \bibinfo {author} {\bibfnamefont {E.}~\bibnamefont {Ahrn{\'e}}}, \bibinfo {author} {\bibfnamefont {B.}~\bibnamefont {Volkmer}}, \bibinfo {author} {\bibfnamefont {L.}~\bibnamefont {Callipo}}, \bibinfo {author} {\bibfnamefont {K.}~\bibnamefont {Knoops}}, \bibinfo {author} {\bibfnamefont {M.}~\bibnamefont {Bauer}}, \bibinfo {author} {\bibfnamefont {R.}~\bibnamefont {Aebersold}},\ and\ \bibinfo {author} {\bibfnamefont {M.}~\bibnamefont {Heinemann}},\ }\bibfield  {title} {\bibinfo {title} {The quantitative and condition-dependent escherichia coli proteome},\ }\href@noop {} {\bibfield  {journal} {\bibinfo  {journal} {Nature biotechnology}\ }\textbf {\bibinfo {volume} {34}},\ \bibinfo {pages} {104} (\bibinfo {year} {2016})}\BibitemShut {NoStop}%
\bibitem [{\citenamefont {Lukk}\ \emph {et~al.}(2010)\citenamefont {Lukk}, \citenamefont {Kapushesky}, \citenamefont {Nikkil{\"a}}, \citenamefont {Parkinson}, \citenamefont {Goncalves}, \citenamefont {Huber}, \citenamefont {Ukkonen},\ and\ \citenamefont {Brazma}}]{lukk2010global}%
  \BibitemOpen
  \bibfield  {author} {\bibinfo {author} {\bibfnamefont {M.}~\bibnamefont {Lukk}}, \bibinfo {author} {\bibfnamefont {M.}~\bibnamefont {Kapushesky}}, \bibinfo {author} {\bibfnamefont {J.}~\bibnamefont {Nikkil{\"a}}}, \bibinfo {author} {\bibfnamefont {H.}~\bibnamefont {Parkinson}}, \bibinfo {author} {\bibfnamefont {A.}~\bibnamefont {Goncalves}}, \bibinfo {author} {\bibfnamefont {W.}~\bibnamefont {Huber}}, \bibinfo {author} {\bibfnamefont {E.}~\bibnamefont {Ukkonen}},\ and\ \bibinfo {author} {\bibfnamefont {A.}~\bibnamefont {Brazma}},\ }\bibfield  {title} {\bibinfo {title} {A global map of human gene expression},\ }\href@noop {} {\bibfield  {journal} {\bibinfo  {journal} {Nature biotechnology}\ }\textbf {\bibinfo {volume} {28}},\ \bibinfo {pages} {322} (\bibinfo {year} {2010})}\BibitemShut {NoStop}%
\bibitem [{\citenamefont {Kinsler}\ \emph {et~al.}(2020)\citenamefont {Kinsler}, \citenamefont {Geiler-Samerotte},\ and\ \citenamefont {Petrov}}]{kinsler2020fitness}%
  \BibitemOpen
  \bibfield  {author} {\bibinfo {author} {\bibfnamefont {G.}~\bibnamefont {Kinsler}}, \bibinfo {author} {\bibfnamefont {K.}~\bibnamefont {Geiler-Samerotte}},\ and\ \bibinfo {author} {\bibfnamefont {D.~A.}\ \bibnamefont {Petrov}},\ }\bibfield  {title} {\bibinfo {title} {Fitness variation across subtle environmental perturbations reveals local modularity and global pleiotropy of adaptation},\ }\href@noop {} {\bibfield  {journal} {\bibinfo  {journal} {Elife}\ }\textbf {\bibinfo {volume} {9}},\ \bibinfo {pages} {e61271} (\bibinfo {year} {2020})}\BibitemShut {NoStop}%
\bibitem [{\citenamefont {Kashtan}\ and\ \citenamefont {Alon}(2005)}]{kashtan2005spontaneous}%
  \BibitemOpen
  \bibfield  {author} {\bibinfo {author} {\bibfnamefont {N.}~\bibnamefont {Kashtan}}\ and\ \bibinfo {author} {\bibfnamefont {U.}~\bibnamefont {Alon}},\ }\bibfield  {title} {\bibinfo {title} {Spontaneous evolution of modularity and network motifs},\ }\href@noop {} {\bibfield  {journal} {\bibinfo  {journal} {Proceedings of the National Academy of Sciences}\ }\textbf {\bibinfo {volume} {102}},\ \bibinfo {pages} {13773} (\bibinfo {year} {2005})}\BibitemShut {NoStop}%
\bibitem [{\citenamefont {Kaneko}(2024)}]{kaneko2024constructing}%
  \BibitemOpen
  \bibfield  {author} {\bibinfo {author} {\bibfnamefont {K.}~\bibnamefont {Kaneko}},\ }\bibfield  {title} {\bibinfo {title} {Constructing universal phenomenology for biological cellular systems: an idiosyncratic review on evolutionary dimensional reduction},\ }\href@noop {} {\bibfield  {journal} {\bibinfo  {journal} {Journal of Statistical Mechanics: Theory and Experiment}\ }\textbf {\bibinfo {volume} {2024}},\ \bibinfo {pages} {024002} (\bibinfo {year} {2024})}\BibitemShut {NoStop}%
\bibitem [{\citenamefont {Halabi}\ \emph {et~al.}(2009)\citenamefont {Halabi}, \citenamefont {Rivoire}, \citenamefont {Leibler},\ and\ \citenamefont {Ranganathan}}]{halabi2009protein}%
  \BibitemOpen
  \bibfield  {author} {\bibinfo {author} {\bibfnamefont {N.}~\bibnamefont {Halabi}}, \bibinfo {author} {\bibfnamefont {O.}~\bibnamefont {Rivoire}}, \bibinfo {author} {\bibfnamefont {S.}~\bibnamefont {Leibler}},\ and\ \bibinfo {author} {\bibfnamefont {R.}~\bibnamefont {Ranganathan}},\ }\bibfield  {title} {\bibinfo {title} {Protein sectors: evolutionary units of three-dimensional structure},\ }\href@noop {} {\bibfield  {journal} {\bibinfo  {journal} {Cell}\ }\textbf {\bibinfo {volume} {138}},\ \bibinfo {pages} {774} (\bibinfo {year} {2009})}\BibitemShut {NoStop}%
\bibitem [{\citenamefont {Masel}(2006)}]{masel2006cryptic}%
  \BibitemOpen
  \bibfield  {author} {\bibinfo {author} {\bibfnamefont {J.}~\bibnamefont {Masel}},\ }\bibfield  {title} {\bibinfo {title} {Cryptic genetic variation is enriched for potential adaptations},\ }\href@noop {} {\bibfield  {journal} {\bibinfo  {journal} {Genetics}\ }\textbf {\bibinfo {volume} {172}},\ \bibinfo {pages} {1985} (\bibinfo {year} {2006})}\BibitemShut {NoStop}%
\end{thebibliography}%

\end{document}